\documentclass[iop,twocolappendix]{emulateapj}

\shorttitle{Inner disk radius with flux in Serpens X-1}
\shortauthors{Chiang et al.}

\begin{document}
\title{On the evolution of the inner disk radius with flux in the neutron star low-mass X-ray binary Serpens X-1}
\author{Chia-Ying Chiang\altaffilmark{1}, Robert A. Morgan\altaffilmark{1}, Edward M. Cackett\altaffilmark{1}, Jon M. Miller\altaffilmark{2}, Sudip Bhattacharyya\altaffilmark{3}, and Tod E. Strohmayer\altaffilmark{4}}
\affil{$^1$Department of Physics and Astronomy, Wayne State University,
    666 W. Hancock, Detroit, MI 48202, USA}
\affil{$^2$Department of Astronomy, The University of Michigan, 500 Church Street, Ann Arbor, MI 48109-1046, USA}
\affil{$^3$Department of Astronomy and Astrophysics, Tata Institute of Fundamental Research, Mumbai 400005, India}
\affil{$^4$X-Ray Astrophysics Lab, Astrophysics Science Division, NASA's Goddard Space Flight Center, Greenbelt, MD 20771, USA}

\begin{abstract}

We analyze the latest \emph{Suzaku} observation of the bright neutron star low-mass X-ray binary Serpens X-1
taken in 2013 October and 2014 April. The observation was taken using the burst mode and only 
suffered mild pile-up effects. A broad iron line is clearly detected in the X-ray spectrum. We test different 
models and find that the iron line is asymmetric and best interpreted by relativistic reflection. The 
relativistically broadened iron line is generally believed to originate from the innermost regions of 
the accretion disk, where strong gravity causes a series of special and general relativistic effects. 
The iron line profile indicates an inner radius of $\sim8$ $R_{\rm G}$, which gives an upper
limit on the size of the neutron star. The asymmetric iron line 
has been observed in a number of previous observations, which gives several inner radius 
measurements at different flux states. We find that the inner radius of Serpens X-1 does not 
evolve significantly over the range of $L/L_{\rm Edd}\sim0.4-0.6$, and the lack of flux dependence 
of the inner radius implies that the accretion disk may be truncated outside the innermost stable circular orbit
by the boundary layer rather than the stellar magnetic field.

\end{abstract}


\section{Introduction}

A low-mass X-ray binary (LMXB) is a compact system that is composed of a 
stellar-mass black hole or a neutron star with a low-mass ($\lesssim$ 1 $M_{\odot}$) companion. 
Both black hole (BH) and neutron star (NS) LMXBs are known to exhibit a number of spectral 
states, and show different behavior on color-color or hardness-intensity diagrams. 
\citet{Hasinger89}, using a number of observations taken by the European X-ray 
Observatory ({\it EXOSAT}), classified NS LMXBs as two types, ``atoll" and 
``Z" sources, based on their X-ray luminosity, spectral and timing properties. The Z 
sources show three-branches (horizontal, normal and flaring branches), Z-shaped 
color-color diagrams and radiate at luminosities close to the Eddington luminosity 
($L_{\rm EDD}$).  The atoll sources display fragmented color-color diagrams with the
island state appearing isolated from the so-called banana branch,
and cover a larger luminosity range. The two classes of sources 
differ in their spectral and timing properties. The Z sources usually go through their tracks 
on the color-color diagram in only one day, whereas the atoll sources show state transitions on 
longer timescales (days to weeks). The X-ray spectra of Z sources are ``soft" on all 
branches, and those of atoll sources are ``soft" at high luminosities and ``hard" at
low luminosities.  \citet{Lin09} studied the outburst and decay of XTE J1701-462 and 
found the object showed all characteristics of Z and atoll sources when its luminosity 
decreases from super-Eddington values toward low Eddington fractions. This implies
that whether a neutron star is an atoll or a Z is determined by the Eddington fraction.

It is generally believed that a geometrically thin, optically thick accretion disk \citep{SS73} 
is formed around the central source when accreting matter from the companion star. The 
accretion disk emits thermal emission which can be described as a quasi-blackbody
component. Seed photons from the accretion disk can be Compton up-scattered to high
energies, which can be approximated as a power-law component. In NS LMXBs,
additional high-energy photons coming from the boundary layer originating from 
the hot flow between the accretion disk and the neutron star surface may be present 
in the X-ray spectrum as well \citep{Barret01,Lin07}. High-energy photons from 
either the Comptonized, power-law component or the boundary layer emission can be 
absorbed by the accretion disk, and then atomic transitions take place to cause 
``reflected" fluorescent lines in the X-ray spectrum \citep{Guilbert88,Lightman88}. 
Reflection coming from the innermost area of the accretion disk is likely shaped by 
relativistic effects due to strong gravity of the compact central object 
\citep{diskline,Fabian00,Miller07,Fabian10}. The most prominent feature in the 
relativistic reflection spectrum is the skewed Fe K line, which has been widely observed in BH 
\citep[e.g.,][]{Miller07,Reis09,Reis10} and NS LMXBs \citep[e.g.,][]{Bha07,Cackett08,Cackett10}. 
By measuring the shape of the Fe K line, a reliable method to determine the inner radius 
of the accretion disk has been developed.

During the high-flux, soft states, in which the X-ray spectra of LMXBs are 
dominated by thermal emission, the accretion disk extends close to the compact object 
and those of BH LMXBs reach the innermost stable circular orbit (ISCO), while those of
NS LMXBs may not because of truncation caused by stellar surface or magnetic fields. 
When sources evolve down to hard states, in which a power-law-like, 
Comptonized component dominates the X-ray spectrum, it is thought that the 
accretion disk recedes and geometrically thick, optically thin Advection-Dominated 
Accretion Flows (ADAF) replace the inner accretion disk \citep{Shapiro76,Narayan95}. 
The theory suggests that state transitions are results of changes in the innermost 
extent of the accretion disk and predicts that the accretion disk is truncated during
hard states (see e.g., the review by \citealt{DGK07}). Nonetheless, relativistically 
blurred iron lines have been observed in hard states of both BH \citep{Miller06,Reis09,Reis10} 
and NS LMXBs \citep{Degenaar15,DiSalvo15,Ludlam16}, which indicates that the accretion disk 
is not truncated at large radii in the hard states above around 1\% $L/L_{\rm Edd}$
and leads to challenges for the ADAF scenario.  Exactly at what flux disk truncation occurs
remains unclear.  However, one clear example of disk truncation was observed by \citet{Tomsick09} 
in the BH LMXB GX~339$-$4, who found a narrower iron line at $\sim$ 0.14\% $L/L_{\rm Edd}$, 
implying an inner radius truncated at a large radius of $>35$ $R_{\rm G}$.

 \citet{Cackett10} analyzed iron lines in a large sample
of NS LMXBs and found no obvious dependence of the inner radius on flux, though
some sources in the sample were only observed once. A different approach is to study a single 
source with many observations to look for changes around the color-color diagram. 4U~1636$-$53 is one
of the most observed NS LMXBs, and shows a broad iron line \citep{Pandel08,Cackett10,Sanna13,Sanna14}.
However, the iron line does not show a clear evolution around the color-color diagram \citep{Sanna14}.
What drives the evolution of color-color/hardness-intensity diagrams still remains unknown \citep[e.g.,][]{Homan10}. 
Probing the evolution of the inner radius is an approach to test theory and helps to 
better understand transitions between states. In order to tackle this issue, multiple 
observations at different flux states on a single source are preferred.

Serpens X-1 was classified as an atoll source and  observed with all major X-ray missions 
in the past. In many of the observations relativistic iron lines have been reported 
\citep{Bha07,Cackett08,Cackett10,Miller13,Chiang16}, making it an ideal 
source for probing the evolution
of the inner radius of NS LMXBs. In the present work we analyze the latest and longest
\emph{Suzaku} observation, and further study the flux dependence of the inner radius 
by comparing with archival observations.
All spectral fitting was performed using the XSPEC 12.8.2 package \citep{Arnaud96} with ``wilm" 
abundances \citep{Wilms00}. Uncertainties are quoted at 90\% of confidence level
in this paper if not stated in particular.

\section{Data Reduction} 

\emph{Suzaku} observed Serpens X-1 during 2013 October 1, 2014 March 13 and April 10 
(obs. ID 408033010, 408033020 and 408033030), resulting in $\sim$ 130 ks, 
$\sim$ 82 ks and $\sim$ 23 ks exposure time (HXD/PIN detector), respectively.
The X-ray Imaging 
Spectrometer (XIS) detectors were operated in burst mode, in order to limit the exposure 
time per frame and to reduce the effects of pile-up. XIS0 and XIS1 were operated in the 
1/4-window mode and 0.135 s frame time, and XIS3 in the full-window mode and 0.1 s 
frame time. A readout streak is observable in the full-window mode, and can be used 
to estimate the effect of out of time events in the 1/4-window burst mode observations. Both $3\times3$ 
and $5\times5$ viewing modes were used in these observations. The data were reduced 
following the \emph{Suzaku} Data Reduction Guide. Event files were combined
for each viewing mode and spectra were extracted from the combined event files.
An overall spectrum of each detector was produced by combining the $3\times3$ and 
$5\times5$ spectra using the ADDASCASPEC tool.
Since Serpens X-1 is a fairly bright 
source ($>130$ counts s$^{-1}$), the observation is likely piled-up even in burst mode. We 
filtered out events that exceed the XIS telemetry limits and proceeded to estimate pile-up 
effects. 

\begin{figure}
\begin{center}
\includegraphics[scale=0.53,trim={2.8cm 11cm 2.8cm 1.8cm}]{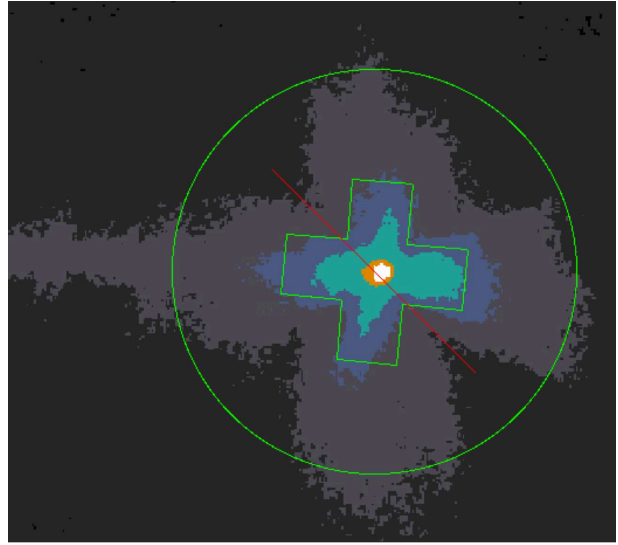}
\caption{An image from the XIS3 detector. The outer green circle is the source region with
a radius of 100$\arcsec$, and the inner plus sign is the exclusion region which is used to correct pile-up effects. A readout
streak can be seen on the figure.}
\label{detector}
\end{center}
\end{figure}

Pile-up effects can be corrected using two methods. We use a concentric circle with
a outer radius of 100$\arcsec$ as the source region. 
The coordinates of the region center are [$\alpha=18^{\rm h}39^{\rm m}56.201^{\rm s}$, $\delta=+5^{\circ}02'04.58''$]
for 408033010, [$\alpha=18^{\rm h}39^{\rm m}57.397^{\rm s}$, $\delta=+5^{\circ}02'07.63''$] for 408033020 and 
[$\alpha=18^{\rm h}39^{\rm m}58.513^{\rm s}$, $\delta=+5^{\circ}02'13.88''$] for 408033030.
We then extracted a series of 
source spectra using an annulus with this outer radius, and varying the inner radius. 
The spectra were fitted in XSPEC, and by comparing the parameters of each spectrum, we 
found that the spectral shapes and fitting parameters began to settle when the radius of 
the inner exclusion region is 20$\arcsec$-30$\arcsec$ or larger, which is consistent with 
previous observations \citep{Cackett10}. The other way to do pile-up correction is to use 
a script in the Interactive Spectral Interpretation System (ISIS) which provides a visual 
presentation of areas of pile-up in the clean event files
\footnote{http://space.mit.edu/ASC/software/suzaku/pile\_estimate.sl}.
Based on the shapes of the areas with a high pile-up fraction ($\geq0.05$), we found 
that the areas affected by pile-up were not circularly-distributed about but rather in the 
shape of a Maltese cross, which is the shape of the point spread function (PSF) for the 
telescope. This prompted the use of a plus-sign shaped exclusion region where each
side of the polygon is 30$\arcsec$ in length with right angles (see Figure \ref{detector}), instead of
a circular region, to correct pile-up effects, as a circular exclusion region would filter
more photons than actually needed. By comparing the pre-correction and 
post-correction spectra, we find that pile-up effects in this observation are fairly mild.
Pile-up only caused slight differences between $\sim1-4$ keV and above $\sim$ 7 keV 
in the spectra and did not affect the iron line profile.

The readout streak was present in the data from XIS3 (see Figure \ref{detector}), 
and thus we did not combine the spectrum of XIS3 with that of the other
front-illuminated detector, XIS0. We measured
the counts taken from a region on the readout streak away from the source and found
that on average $\sim$ 12\% of events in the source region were caused by the readout
streak in the full-window mode observation, implying $\sim$ 9\% of affected photons
in the 1/4-window mode observations. There is no significant variability between
observations, and the XIS0, XIS1, and XIS3 spectra shown in the following were 
combined from the three observations.

The Hard X-ray Detector (HXD) was operated as well, in XIS nominal pointing mode, to 
collect high-energy photons of the source. We reduced the data following standard 
procedures. The background spectrum was produced by combining the non-X-ray 
background and the cosmic-X-ray background. As the background dominates above 
$\sim26.0$ keV, we only use the data over the 15.0-26.0 keV energy range in
the following analyses. For both XIS and HXD/PIN detectors, we used the ADDASCASPEC
tool to combine the spectrum from each observation and obtained a total spectrum for
each detector.

\section{Data Analysis}

\subsection{Continuum Model}

Serpens X-1 has been observed by different instruments at different times. There
are several continuum models used in previous work, using different combinations of a disk
blackbody, a blackbody and a power-law. A continuum model consisting 
of all three components was used most often, 
especially in broadband observations (2006 \emph{Suzaku} observation:  
\citealt{Cackett08,Cackett10}; 2013 \emph{NuSTAR} observation: \citealt{Miller13}).
Continuum models including only a disk blackbody and a blackbody, or a blackbody 
and a power-law were used in previous observations as well, but only in observations
with limited energy ranges \citep{Bha07,Chiang16}. We test the continuum 
models which successfully fit the archival data on our latest \emph{Suzaku} 
observation.

We use the XIS data over the $1-10$ keV energy band, and the PIN spectrum over 
$15-26$ keV. The $1.45-2.5$ keV in the XIS spectra were ignored due to poor 
calibration around the Si and Au edges. An iron K emission line is detected in the spectra.
In order not to bias continuum measurements, we first excluded the $5.0-7.5$ keV 
energy band (the iron line band) to test various continuum models in the initial fits.
The Galactic absorption is 
modeled using TBABS in XSPEC; the accretion disk emission is modeled with
DISKBB; the blackbody component which is likely
caused by the boundary layer is accounted for using BBODY. We added a constant
between the spectra (with that of the XIS0 spectrum frozen at 1.00)
and linked all parameters except normalizations of DISKBB and BBODY. As there are 
mild spectral differences between the spectra of different XIS detectors, which are 
caused by calibration and different observation modes, the normalizations of 
DISKBB and BBODY components are free to vary for better results. The continuum
model with all three components results in a $\chi^2/$d.o.f. of 2617/2323, and other models yield worse fits. 
The result is consistent with previous work using broadband X-ray data. We also test thermal Comptonization
model using NTHCOMP \citep{nthcomp96,nthcomp99} in XSPEC, and it does not 
improve the fit. Since the TBABS*(DISKBB + BBODY + POWERLAW) continuum 
fits significantly better than other continuum models, we use it in all following analyses.

The best-fitting continuum model parameters are listed in Table \ref{continuum}.
It can be seen that the neutral absorption column density $N_{\rm H}$ is higher 
than any value previously reported. The value of $N_{\rm H}$ reported in previous 
literature spans a wide range around $\sim$ (3-7.5)$\times10^{21}$ cm$^{-2}$ 
\citep{Cackett08,Cackett10,Ng10}, and those obtained from \emph{Suzaku} data 
are higher than the others. The highest $N_{\rm H}$ obtained from past 
work was $(7.5\pm0.2)\times10^{21}$ cm$^{-2}$ (1$\sigma$ uncertainty) obtained 
from pervious \emph{Suzaku} observation \citep{Cackett10}, while here 
$8.6^{+0.9}_{-0.7}\times10^{21}$ cm$^{-2}$ is required to fit the data though
the same absorption model is used. However, the abundance table used in 
\citet{Cackett10} was \citet{AG89}. If we use the \citet{AG89} abundance to fit
the data, we obtain a $N_{\rm H}$ of $5.9^{+0.5}_{-0.3}\times10^{21}$ cm$^{-2}$, which
is lower than the value \citet{Cackett10} reported and consistent with
the historical range.

Note that the constant for the XIS3 is $1.25^{+0.04}_{-0.05}$. 
In previous section we found that the effect caused by the readout streak is $\sim$ 9\%. 
The constant is still slightly higher than expected when the effect is 
accounted for due to unknown reason. The best-fitting values of remaining parameters 
shown in Table \ref{continuum} are consistent with \citet{Cackett08}.

\begin{deluxetable*}{llrrrrr}
\tablecaption{Best-fitting continuum model parameters.}
\tablehead{
Component &Parameter & XIS0 &  XIS1 & XIS3 & PIN}
\startdata
Constant &  & (1.00) & $0.97^{+0.02}_{-0.01}$ & $1.25^{+0.04}_{-0.05}$ & $0.65^{+0.03}_{-0.01}$\\
TBABS & $N_{\rm H}$ ($10^{22}$ cm$^{-2}$)& $0.86^{+0.09}_{-0.07}$ & \nodata & \nodata &\nodata\\
DISKBB & $kT_{\rm disk}$ (keV)& $1.36\pm0.01$ & \nodata & \nodata &\nodata\\
 & $N_{\rm disk}$ & $75\pm4$ & $71^{+5}_{-4}$ & $79\pm5$ &\nodata\\
BBODY & $kT_{\rm bb}$ (keV)& $2.29^{+0.02}_{-0.03}$ & \nodata & \nodata &\nodata \\
 & $N_{\rm bb}$ ($10^{-2}$)& $3.6\pm0.1$ & $3.3\pm0.1$ & $3.3\pm0.1$ &\nodata \\
 POWERLAW & $\Gamma$ & $3.44^{+0.35}_{-0.33}$  & \nodata & \nodata &\nodata\\
 & $N_{\rm pow}$ & $0.71^{+0.22}_{-0.16}$ & \nodata & \nodata &\nodata
\enddata
\label{continuum}
\tablecomments{ Parameters of the HXD/PIN
spectrum are bound to those of the XIS0 spectrum. We use XIS data over the
$1-10$ keV ($1.45-2.5$ keV excluded due to calibration issues), and PIN data over 
the $15-26$ keV energy bands. The $5-7.5$ keV (iron line energy band) was excluded
here. The model results in a $\chi^2/$dof of 2617/2323.}
\end{deluxetable*}

\subsection{Phenomenological Models} \label{sec:phe}

\begin{figure}
\begin{center}
\includegraphics[scale=0.5]{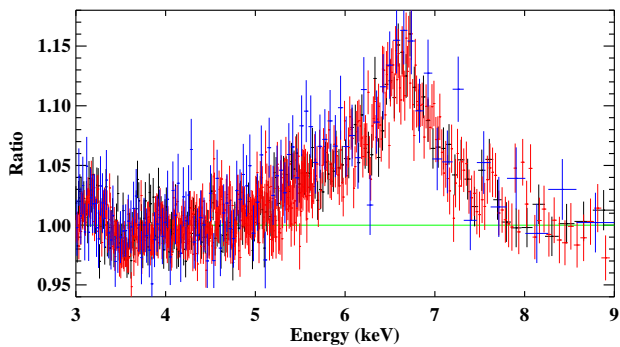}
\caption{Ratio of the data to the continuum model, showing the iron line profile. The continuum model was fit to the spectra with the $5.0-7.5$ keV region ignored. The XIS0, XIS1, and XIS3 data are presented
by black, red and blue points, respectively. An broad iron emission line is clearly present
in the spectrum.}
\label{profile_continuum}
\end{center}
\end{figure}

After fitting the continuum we reload the $5.0-7.5$ keV data and find 
that an iron emission line is clearly detected (see Fig. \ref{profile_continuum}). 
In order to model the broadband X-ray spectrum with the iron line, we add a 
Gaussian component on top of the continuum model.
We restrict the
energy of the Gaussian line to be $6.4-6.97$ keV, which is the energy range
that the Fe K$\alpha$ line can appear in the spectrum. The fitting results
are listed in Table \ref{phe}. The absorption column density is slightly 
higher than the value of continuum fitting. The photon index $\Gamma$
tends to be higher than those previously reported. Note that the Gaussian 
width $\sigma=0.74^{+0.04}_{-0.05}$ keV is higher than the broadest value 
in published literature ($\sigma$ = 0.61 keV, see \citealt{Cackett12}), 
indicating the existence of a broad Fe K line. In addition, the equivalent 
width (EW) of the line $192^{+17}_{-15}$ eV is larger than values 
obtained from previous Serpens X-1 observations 
\citep{Bha07,Cackett08,Cackett10,Chiang16}. These may 
imply that the Gaussian component is fitting part of the continuum, and 
whether the Fe K line is symmetric or not should be further investigated.

\begin{deluxetable*}{llrrrr}
\tablecaption{Spectral fit parameters for the phenomenological models}
\tablehead{
Component &Parameter & Gaussian &  DISKLINE & RELLINE}
\startdata
TBABS & $N_{\rm H}$ ($10^{22}$ cm$^{-2}$)& $0.97^{+0.07}_{-0.08}$ & $0.81^{+0.10}_{-0.05}$ & $0.82^{+0.09}_{-0.06}$\\
DISKBB & $kT_{\rm disk}$ (keV)& $1.35^{+0.02}_{-0.01}$ & $1.35\pm0.01$ & $1.35\pm0.01$ \\
 & $N_{\rm disk}$ & $78^{+5}_{-4}$ & $75^{+13}_{-3}$ & $75^{+4}_{-3}$ \\
BBODY & $kT_{\rm bb}$ (keV)& $2.31\pm0.01$ & $2.27\pm0.03$ & $2.27\pm0.03$ \\
 & $N_{\rm bb}$ ($10^{-2}$)& $3.6\pm0.1$ & $3.7\pm0.1$ & $3.7\pm0.1$ \\
 POWERLAW & $\Gamma$ & $3.87^{+0.23}_{-0.28}$  & $3.22^{+0.42}_{-0.22}$ & $3.24^{+0.41}_{-0.24}$\\
 & $N_{\rm pow}$ & $0.98^{+0.21}_{-0.19}$ & $0.60^{+0.21}_{-0.11}$ & $0.61^{+0.21}_{-0.11}$\\
\tableline
Gaussian & $E_{\rm line}$ & $6.4^{+0.02}$ & \nodata & \nodata \\
 & $\sigma$ (keV) & $0.74^{+0.04}_{-0.05}$ & \nodata & \nodata \\
 & $N_{\rm gau}$ ($10^{-3}$) & $10.0\pm0.9$ &\nodata & \nodata \\ 
\tableline
LINE & $E_{\rm line}$ (keV) & \nodata & $6.97_{-0.02}$ & $6.97_{-0.02}$\\
 & $q$ & \nodata & $4.24^{+0.24}_{-0.22}$ & $4.22^{+0.25}_{-0.22}$ \\
 & $R_{\rm in}$ ($GM/c^2$) & \nodata & $8.1^{+0.4}_{-1.2}$ & $8.1^{+0.5}_{-0.4}$ \\
 & $i$ (deg) & \nodata & $22\pm1$ & $23\pm1$\\
 & $N_{\rm line}$ ($10^{-3}$) & \nodata & $7.3^{+1.9}_{-0.5}$ & $7.4^{+0.4}_{-0.6}$\\
 \tableline
EW (eV) & & $192^{+17}_{-15}$  & $147^{+9}_{-12}$ & $146^{+8}_{-12}$\\
 \tableline
$\chi^2/$d.o.f. & & 3763/3352 & 3730/3350 & 3728/3350
\enddata
\label{phe}
\tablecomments{For simplicity, we only present results of XIS1 (the $\chi^2/$d.o.f. is of the entire data set). The constant
of each spectrum remains the same as those displayed in Table \ref{continuum}. Note 
that the emissivity index given by DISKLINE is $\beta$, which is equal to $-q$. We list $q$
here for easier comparison with other models.}
\end{deluxetable*}

\begin{figure}
\begin{center}
\includegraphics[scale=0.5]{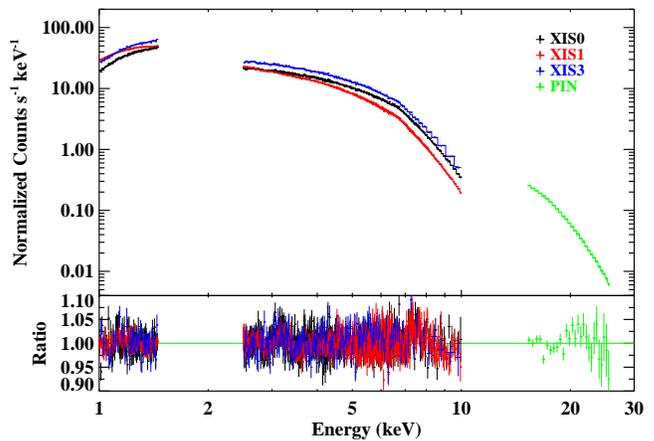}
\caption{The figure shows the XIS0 (black), XIS1 (red), XIS3 (blue) and 
PIN (green) spectra of Serpens X-1 fitted with the continuum plus a 
DISKLINE component. The lower panel is the data/model ratio, and it
can be seen that the model fits the broadband spectrum well.
}
\label{plldra}
\end{center}
\end{figure}

We replaced the Gaussian component with the DISKLINE model to 
fit the spectrum in order to examine if the line profile is better 
interpreted by relativistic effects, and to directly compare results with those previously published in the
literature. It is generally believed that the Fe K line
originates from an accretion disk around objects of strong gravity 
(i.e., black holes or neutron stars) and is shaped by a series of relativistic
effects, including relativistic beaming and gravitational redshift, and 
appear to be broad and asymmetric. We froze the outer disk radius
$R_{\rm out}$ at 1000 $R_{\rm G}$ in the fitting with other parameters
of DISKLINE free to vary. The lower limit of the inner radius
$R_{\rm in}$ in the model is 6 $R_{\rm G}$ and we restricted the inclination angle to
vary between $5-80^{\circ}$. As shown in Table \ref{phe}, the fit improves
by $\Delta\chi^2=33$ with two lower degrees of freedom compared to
the Gaussian model, and the model fits the data well (see Fig. \ref{plldra}). 
The F-test probability is $\sim3.91\times10^{-7}$,
which is convincing that the DISKLINE model explains the data better.
The EW of the Fe line obtained via the DISKLINE model is 
$147^{+9}_{-12}$ eV, which is also consistent with previous results 
\citep{Cackett08,Chiang16}.

The temperatures of the DISKBB and BBODY components are nearly
identical in both models. Although the photon indices $\Gamma$ 
obtained from the models seem to be different, the fluxes of the powerlaw
component are comparable. The differences between $\Gamma$ may
be led by the mild differences in the blackbody temperatures. The 
DISKLINE model gives a slightly lower $N_{\rm H}$, indicating the parameter
may be mildly model-dependent. 

\begin{figure}
\centering
\includegraphics[scale=0.5]{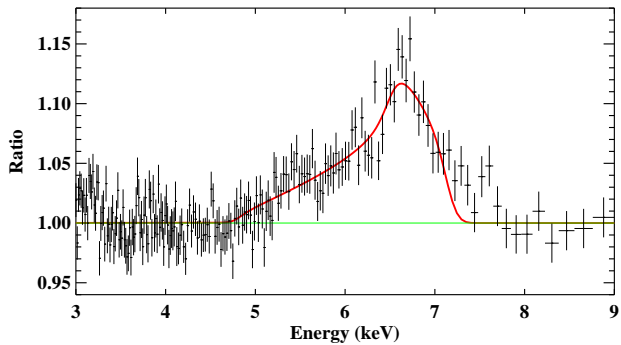}
\caption{The figure shows the iron line profile (XIS0 data only) fitted
with the DISKLINE model.}
\label{profile}
\end{figure}

The parameters of the DISKLINE component are remarkably similar to 
previously reported numbers \citep{Cackett10,Chiang16}.
We obtained an inner radius $R_{\rm in}=8.1^{+0.4}_{-1.2}$ $R_{\rm G}$, 
which is close to a number of previous results
\citep{Cackett12,Miller13,Chiang16}, and the line profile is similar as well 
(see Fig. \ref{profile}). 

We also replaced the DISKLINE 
model with the more modern RELLINE \citep{Dauser10,Dauser13} model 
 (setting the spin parameter a* = 0), and find that 
a best-fitting inner radius of $8.1^{+0.5}_{-0.4}$ $R_{\rm G}$
(see Table \ref{phe}), consistent with the DISKLINE fits. A low inclination angle
$i$ $\sim23\pm1$ is required to fit the data, which is expected as low 
inclination has been reported in a series of past work 
\citep{Cackett10,Cackett12,Miller13,Chiang16}. Optical observations of
Serpens X-1 also point to a low inclination of less than 10$^\circ$
\citep{Cornelisse13}, which is in good agreement with the result
of X-ray observations. Note that it is possible that the inner disk 
can be slightly mis-aligned with the binary orbit, for instance, see the case of GRO~J1655$-$40
\citep{Hjellming95,Greene01}.

\subsection{Reflection Model}

We learn from the tests of phenomenological models that the iron line is 
better explained by relativistic reflection. In order to confirm this scenario,
we replace the DISKLINE component with a self-consistent reflection 
model which includes a broadband continuum and all emission lines 
originated from an illuminated accretion disk. Since the blackbody emission 
dominates the flux above 10 keV which implies the illuminating source 
is the boundary layer emission or stellar surface, we use the 
BBREFL \citep{Ballantyne04} grid, which calculates reflected emission from 
an accretion disk illuminated by a blackbody component, to model the 
reflection. We use the convolution kernel RELCONV \citep{Dauser10}
to account for relativistic effects.

\begin{figure}
\begin{center}
\includegraphics[scale=0.5]{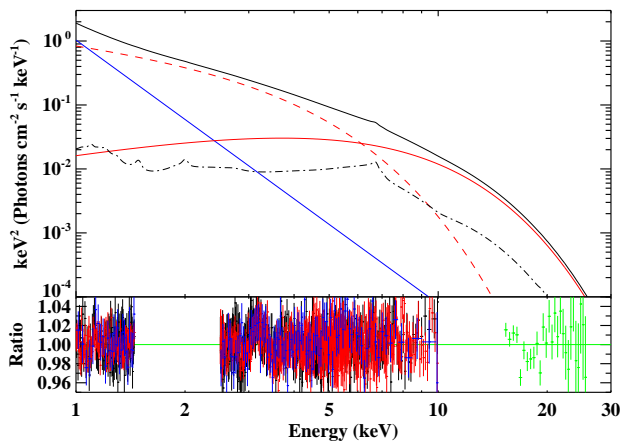}
\caption{The upper panel shows the decomposed best-fitting model. 
The red dash line represents the DISKBB component, the red solid line
the BBODY component, the blue solid line the POWERLAW component, 
and the black dot dash line the blurred reflection continuum 
RELCONV$\times$BBREFL. The black sold line on top is the total model 
with all the components combined. The lower panel shows the data/model 
ratio, in which black, red, blue, and green data points stand for XIS0, XIS1, 
XIS3, and PIN data, respectively.}
\label{model}
\end{center}
\end{figure}

We list best-fitting parameters in Table \ref{reflection}. It can be seen that
the fit has been improved by $\Delta\chi^2=99$ compared to the DISKLINE
model. The decomposed model and data/model ratio can be found in 
Fig. \ref{model}. The values of the inner radius, inclination angle,
and emissivity index are lower than those obtained from the DISKLINE model but 
consistent with the numbers reported in \citet{Miller13}. The reflection model
gives a slightly higher $N_{\rm H}$ than those obtained from phenomenological 
models, but the values are roughly consistent within $3\sigma$. We again obtain 
a low inner radius $R_{\rm in}=6.7^{+0.8}_{-0.7}~R_{\rm G}$, and a low inclination 
angle $i=9^{+7}_{-4}$ degrees, implying measurements in previous work are robust.

\begin{deluxetable}{lrr}
\tablecaption{Spectral fit parameters for the reflection model}
\tablehead{
Component &Parameter & BBREFL}
\startdata
TBABS & $N_{\rm H}$ ($10^{22}$ cm$^{-2}$)& $1.00^{+0.07}_{-0.04}$  \\
DISKBB & $kT_{\rm disk}$ (keV)& $1.29^{+0.01}_{-0.02}$  \\
 & $N_{\rm disk}$ & $95^{+5}_{-6}$  \\
BBODY & $kT_{\rm bb}$ (keV)& $2.33\pm0.01$   \\
 & $N_{\rm bb}$ ($10^{-2}$)& $3.1^{+0.1}_{-0.2}$   \\
 POWERLAW & $\Gamma$ & $4.14^{+0.24}_{-0.20}$   \\
 & $N_{\rm pow}$ & $1.03^{+0.20}_{-0.12}$  \\
\hline
RELCONV & $q$ & $2.6^{+0.1}_{-0.2}$ \\
 & $R_{\rm in}$ ($GM/c^2$)  & $6.7^{+0.8}_{-0.7}$\\
 & $i$ (deg)  & $9^{+7}_{-4}$\\
\hline
BBREFL & log $\xi$ & $3.05^{+0.07}_{-0.06}$ \\
 & $N_{\rm bbrefl}$ ($10^{-26}$) & $1.5^{+0.2}_{-0.1}$\\
\hline 
$\chi^2/$d.o.f. & & 3631/3350 
\enddata
\label{reflection}
\tablecomments{For simplicity, we only
present results of XIS1 (the $\chi^2/$d.o.f. is of the entire data set though). The constant
of each spectrum remains the same as those displayed in Table \ref{continuum}. }
\end{deluxetable}

\section{Discussion}

Serpens X-1 has been observed by different missions in the past, and 
there are several measurements of the inner radius which can be
found in the literature \citep{Bha07,Cackett08,Cackett10,Miller13,Chiang16}.
As these archival observations were taken at 
different times, and possibly in different flux states, these offer the possibility
to probe the evolution of the inner radius in Serpens X-1. We then 
check previous work to find fluxes and inner radii obtained from other
observations, and compare these numbers with those of the present 
work. The $0.5-25.0$ keV absorption-corrected (unabsorbed) fluxes were
converted into luminosities using a distance of 7.7 kpc (from X-ray burst 
properties, see e.g. \citealt{Galloway08}). We list the $0.5-25.0$ keV 
luminosity and inner radius quoted from a number of previous observations 
(see Table \ref{literature}). Note that these numbers were calculated 
using phenomenological models in which the DISKLINE model was 
used to model the iron line. In some references, measurements based
on self-consistent relativistic reflection models are also listed, and we
find most the results are consistent with phenomenological model 
calculations. In \citet{Cackett10} and \citet{Miller13} a continuum 
model which is the same with that of the present work was used, while in 
\citet{Chiang16} a continuum model composed of a blackbody and a
powerlaw components was used, but the authors reported that the line
parameters are not continuum-dependent in that observation.

\begin{deluxetable*}{lrrrrr}
\tablecaption{Luminosity and inner radius $R_{\rm in}$ 
of each previous Serpens X-1 observation}
\tablehead{
Observation  & Luminosity (erg/s) & $L/L_{\rm Edd}$  & Inner Radius $R_{\rm in}$ ($R_{\rm G}$)  & Reference}
\startdata
\emph{Suzaku} (2006) & $(8.5\pm0.1)\times10^{37}$ & 0.48 & $8\pm0.3$ & \citet{Cackett10}\\
\emph{XMM} (2004, Obs ID 0084020401)  & $(5.0\pm0.1)\times10^{37}$ & 0.28 & $25\pm8$ & \citet{Cackett10}\\
\emph{XMM} (2004, Obs ID 0084020501)  & $(3.6\pm0.1)\times10^{37}$ & 0.20 & $14\pm1$ & \citet{Cackett10}\\
\emph{XMM} (2004, Obs ID 0084020601)  & $(4.6\pm0.1)\times10^{37}$ & 0.26 & $26\pm8$ & \citet{Cackett10}\\
\emph{NuSTAR} (2013) & $1.1\times10^{38}$* & 0.62 & $10.6\pm0.6$ & \citet{Miller13}\\
\emph{Chandra} (2014) & $(6.7\pm0.1)\times10^{37}$ & 0.38 & $7.7\pm0.1$ & \citet{Chiang16}\\
\emph{Suzaku} (2013-14) & $(7.1\pm0.4)\times10^{37}$ & 0.40 & $8.1^{+0.4}_{-1.2}$ & present work
\enddata
\tablecomments{ The luminosity was 
calculated based on the $0.5-25.0$ keV absorption-corrected flux 
(measured using a phenomenological model) and
a distance of 7.7 kpc. Luminosities quoted in \citet{Cackett10} have
much larger uncertainties due to including 25\% of error on
distance. In this work we only compare results of the same source,
and uncertainties on distance were excluded.
*Note that the flux of the \emph{NuSTAR}
observation was measured using the $0.5-40.0$ keV band. Since
the high-energy tail of this source is weak, the $25.0-40.0$ keV
energy band only contributes little flux.}
\label{literature}
\end{deluxetable*}

\begin{figure}
\begin{center}
\includegraphics[scale=0.5]{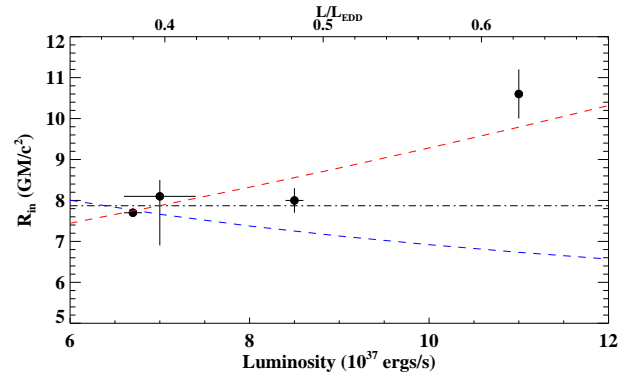}
\caption{The figure is the visualization of Table \ref{literature}
shows the inner radii measured in previous literature
in different flux states. We assume a mass of 1.4$M_{\odot}$ 
to convert the luminosity into the Eddington ratio $L/L_{\rm Edd}$. The black dash
dot line marks the weighted mean of the inner radius measurements; the blue
dash line shows the prediction of the magnetosphere radius; the red dash line
presents the prediction of the boundary layer model.}
\label{rin}
\end{center}
\end{figure}

It can be clearly seen that in Table \ref{literature} that the inner radius spans a 
range between $\sim8-25$ $R_{\rm G}$ and the flux between 
$\sim(3-11)\times10^{37}$ erg s$^{-1}$ ($\sim0.2-0.6$ Eddington ratio 
$L/L_{\rm Edd}$ if assuming a mass of 1.4 $M_{\odot}$).
Although some of the 2004 \emph{XMM-Newton} observations give large 
inner radii of $\sim25$ $R_{\rm G}$, these measurements are 
not well-constrained and analyses by \citet{Bha07} give smaller values. 
Note that the \emph{XMM} data are
the least reliable among these archival data due to short exposure and 
calibration issues of EPIC-PN timing mode data (see e.g. \citealt{Walton12}), 
though the broad line does not seem to suffer from pile-up effects \citep{Miller10}.
In order to see how the inner radius evolves with the flux more clearly, 
we plot the results in Fig. \ref{rin} but exclude \emph{XMM} results and only
present well-constrained measurements for clarity.

The weighted mean of all inner radius measurements is $\sim7.9~R_{\rm G}$, 
which is $\sim$16 km if assuming the neutron star mass to be 1.4 $M_{\odot}$.
The measurement of the inner radius gives an upper limit of the size of the 
neutron star.  It can be seen in Fig. \ref{rin} that more constrained measurements of the 
inner radius fall within a narrow range, implying that the accretion disk is not truncated at
a large radius at fluxes $\ga 0.4$ $L/L_{\rm Edd}$. Furthermore, from Fig. \ref{rin} it seems that
the inner radius does not show a strong flux dependence. For a neutron star, the ISCO 
depends on the mass and radius of the neutron star and its equation of state.  However, 
it should lie somewhere between $5-6$ $R_{\rm G}$ for reasonable parameters \citep{Bha11}. 
Our measured inner disk radius of $8~R_{\rm G}$, is therefore larger than the expected 
ISCO and stellar surface.
The disk could therefore be truncated by either the stellar surface, the boundary layer between 
the disk and star, or the neutron star magnetic field. 

If we assume that the magnetic field truncates the disk and therefore the magnetospheric radius is $7.9~R_{\rm G}$, we can estimate the
magnetic field strength \citep[see e.g.,][]{illarionov75,Ibragimov09,Cackett09}. Given that the unabsorbed 
$2-25$ keV flux of Serpens X-1 in this observation is $F_{\rm 2-25 keV}=6.62\times10^{-9}$
erg cm$^{-2}$ s$^{-1}$, we estimate the magnetic dipole moment $\mu = 1.23\times10^{26}$ 
G cm$^{3}$ following the Equation (1) and assumptions in \citet{Cackett09}.  This corresponds to
a magnetic field strength at the poles of $B = 2.5\times10^8$~G (assuming a 10 km neutron star).

If the magnetic field is truncating the disk, then we would expect the inner disk radius to change with
flux, since the magnetospheric radius depends on the mass accretion rate as $\dot{m}^{-2/7}$.  
To show the expected flux-dependence of the magnetospheric radius, we assume it is equal 
to 8 $R_{\rm G}$ at $L = 6\times10^{37}$ erg s$^{-1}$ and show the relation in Fig. \ref{rin} (see the blue dash line).
The factor of $\sim$1.5 in flux that are covered all observations, would lead to an expected change 
by a factor of 0.9, with the highest flux observation having the smallest inner disk radius. 
This is not what we see here, with the largest radius being at the highest flux. This may be more 
easily explained if the disk is truncated by the boundary layer, with the boundary increasing in size 
with mass accretion rate, as expected from theoretical considerations 
\citep{Popham01}. We also show the dependence of the boundary layer radius with flux in Fig. \ref{rin} 
(red dash line) using equation 25 of \citet{Popham01}, and relating $\dot{M}$ to $L$ through 
$L_{\rm acc} = GM\dot{M}/R_{*}=1.17\times10^{37}~(\dot{M}/10^{-9}~M_{\odot}$ yr$^{-1}$) erg s$^{-1}$, 
where $R_{*}$ is the neutron star radius assumed to be 10 km and the
neutron star mass $M=1.4~M_{\odot}$ (see the red dash line in Fig. \ref{rin}). The increase in 
inner disk radius at the highest mass accretion rate is similar to the size of the boundary layer 
predicted by \citet{Popham01}. The Popham \& Sunyaev boundary layer model, however, 
assumes a nonmagnetic neutron star and the addition of a magnetic field will presumably 
change the boundary layer structure somewhat.

\citet{Steiner10} found values of the inner radius of the black hole binary 
LMC X-3 to be consistent within $4-6$\% across eight X-ray missions,
implying the inner radius is stable over different flux states. In neutron
star LMXBs the evolution of inner radius has also been studied. 
\citet{Lin10} found that the accretion disk of the neutron star system 
4U~1705$-$44 is close to the neutron star during soft states based on 
the broad iron line. In hard states of 4U~1705$-$44, whether 
the accretion disk is truncated is however debated. \citet{DAi10} 
suggested the accretion disk to be truncated, while \citet{DiSalvo15} 
indicated that it is not truncated at large radii, down to 
$\sim$ 3\% of Eddington luminosity. 4U~1636$-$53 also shows no clear evolution in inner disk radius
around the color-color diagram \citep{Sanna14}.
Serpens X-1 has only been observed during soft states, and the inner 
radius evolution during hard states remains unclear. But in soft states
the inner radius is roughly constant over fluxes changing by a factor of 
around 1.5, and the accretion disk stays close in at least down to 
$\sim$0.4$L/L_{\rm Edd}$. 

\section{Conclusion}

We present detailed spectral analysis of the latest, longest \emph{Suzaku}
observation of the neutron star LMXB Serpens X-1 in this work. The 
continuum is best explained by the combination of a disk blackbody 
component, a single-temperature blackbody component and a powerlaw
component, which is consistent with the results of previous broadband 
observations. We find that the relativistic reflection scenario is the better
interpretation of the iron line profile shown in this observation. The line
parameters obtained are consistent with a number of previous work.
By comparing the inner radius obtained from different observations taken
at different flux states, we find no strong evidence that the inner radius
evolves with flux, with the inner radius staying at $\sim$8 $R_{\rm G}$.
The observation with the highest flux shows the largest inner disk radius, 
implying that a boundary layer, rather than the stellar magnetic field, truncates
the disk outside the ISCO.
Results of current data indicate that the inner radius stays
unchanged at soft states. To further confirm if this stands true for hard
states, multiple observations at different times on the same source should
be performed to cover a wider range of flux states.

\acknowledgments
This work was greatly expedited thanks to the help of Jeremy Sanders in
optimizing the various convolution models. We thank Masahiro Tsujimoto
for help choosing the optimal XIS observing modes.
C.Y.C. and E.M.C. gratefully 
acknowledge support provided by NASA through Chandra Award Number 
GO4-15041X issued by the Chandra X-ray Observatory Center, which is 
operated by the Smithsonian Astrophysical Observatory for and on behalf 
of NASA under contract NAS8-03060. R.M. acknowledges support from the 
NSF through a Research Experience for Undergraduates program at Wayne
State University (NSF grant number PHY1460853).

\bibliographystyle{apj}
\bibliography{serx1}

\end{document}